\documentclass[final,5p,times,twocolumn]{elsarticle}

\usepackage{amssymb}
\usepackage{xcolor}
\usepackage{slashed}
\usepackage{cancel}
\usepackage{hyperref}
\usepackage{soul}
\usepackage{mathtools, cuted}

\journal{Physics Letters B}

\begin{document}

\begin{frontmatter}

\title{Q-ball polarization - a smooth path to oscillons}

\author[a]{F. Blaschke}
\affiliation[a]{organization={Research Center for Theoretical Physics and Astrophysics, Institute of Physics, Silesian University in Opava}, 
            addressline={Bezrucovo namesti~1150/13}, 
            city={Opava}, 
            postcode={746 01}, 
            country={Czech Republic}}

\author[b]{T. Romanczukiewicz}
\affiliation[b]{organization={Institute of Theoretical Physics, Jagiellonian University},
            addressline={Lojasiewicza 11}, 
            city={Krakow},
            postcode={30-348},
            country={Poland}}

\author[b]{K. Slawinska}

\author[b,c]{A. Wereszczynski}
\affiliation[c]{organization={International Institute for Sustainability with Knotted Chiral Meta Matter (WPI-SKCM2)},
            addressline={Hiroshima University}, 
            city={Higashi-Hiroshima},
            postcode={739-8526}, 
            country={Japan}}
            
\begin{abstract}
We show that in the complex $\phi^6$ theory the oscillon, together with its spectral structure and the amplitude modulation, arises from the exited Q-ball carrying the bound and the quasi-normal modes.
\end{abstract}


\end{frontmatter}

%%%%%%%%%%%%%%%%%%%%%%%%%%%%%%%%%%%%%%%%%%%%%%%%
\section{Motivation}
\label{motivation}
%%%%%%%%%%%%%%%%%%%%%%%%%%%%%%%%%%%%%%%%%%%%%%%%
Oscillons and Q-balls are examples of non-topological, solitonic-like excitations. At first glance, they are rather different objects. The Q balls are stationary solutions that periodically oscillate in the target complex space \cite{Coleman:1985ki, Friedberg:1976me}. They are stable due to the $U(1)$ Noether charge. Oscillons, on the contrary, are quasi-periodic solutions for which the energy density oscillates in time \cite{Bogolyubsky:1976nx, Gleiser:1993pt, Copeland:1995fq}. The fundamental frequency of the oscillations is, however, accompanied by a tower of frequencies. Oscillons exist even in real field theories and do not carry any charge. Thus, their existence is not related, in an obvious manner, to any conservation law. In fact, they are not ultimately stable, but eventually decay, sometimes extremely slowly \cite{Graham:2006xs, Salmi:2012ta, Zhang:2020bec, Olle:2020qqy}.

Despite these differences, oscillons and Q-balls are quite similar. Both are non-topological, oscillating, and localized excitations requiring an attractive nonlinear potential, see \cite{Zhou:2024mea} for a review. Therefore, for a long time, it has been conjectured that they should be closely related to each other. 

The first example of such a relation was formulated in terms of the I-balls \cite{Kasuya:2002zs}. Here, the stability of oscillons was connected to the quasi-conservation of adiabatic invariants emerging from the $U(1)$ current. 

Recently, another type of oscillon--Q-ball relation has been discovered \cite{Blaschke:2024dlt}. Here, real field oscillons are generated from Q-balls being solutions of some hidden complex field theories. Importantly, at the leading order of the nonlinearity, these theories are universal, that is, oscillons in many real scalar models arise from the same Q-ball. This framework, in a natural way, explains modulations of excited oscillons in (1+1) dimensions as an effect of the formation of the two-oscillon bound state based on a pertinent two Q-ball solution. 

In the present work, we would like to further study the relation between oscillons and Q-balls although in a different context, that is, in a model in which {\it both excitations exist simultaneously}. This happens in complex scalar field theories. Such theories may support Q-balls of the usual form
\begin{equation}
    \phi(x,t)=f_\omega(x)e^{i\omega t}\,, \label{Qball}
\end{equation}
with $\omega <m$, where $\omega$ is the internal frequency and $m$ is the mass threshold. Often, there is also a lower bound for $\omega$. Next, purely real, small amplitude fundamental (i.e., without modulations)  oscillons are approximated by
\begin{equation}
    \phi(x,t)\approx \sum_{k=0}^\infty g_n(x)\cos(n\omega_O t)\approx g_0 + g_1(x) \cos ( \omega_O t ). \label{oscillon}
\end{equation}
This expansion is not accurate because oscillators radiate, and one should include radiative tails. Moreover, the frequency $\omega_O$ is not constant in time, nor are the coefficients $g_n(x)$. However, the time scale over which the frequency changes is much larger than the frequency itself. Therefore, such an approximation is usually good enough on a short time scale. The most dominant contribution comes from $g_1(x)$. Here, $g_0=0$ is the vacuum around which the oscillon oscillates.

Typically, the profile functions $f_\omega$ and $g_1$ are quite similar with identical asymptotic behavior. This justifies approximation of the oscillon as an on-top superposition of two Q-balls with opposite charges
\begin{equation}
g_1(x)\cos(\omega t) \approx \frac{1}{2}e^{i\omega t}f_\omega(x) + \frac{1}{2}e^{-i\omega t}f_\omega(x)\,. 
\end{equation}

In this approximated picture, the oscillon and Q-balls are simply differently "target space-polarized" objects. The circularly polarized solution, Q-ball, has a constant energy density, while the linear polarization of the oscillon enforces periodic density oscillations. From this point of view, these objects are (approximately) dual to each other, and each of them can be formed as a superposition of the others. 

In the following analysis, we will exploit this approach and consider a family of 
configurations smoothly connecting the Q-ball and the oscillon
\begin{equation}
    \phi(x,t) = \left(1-\frac{A}{2}\right)e^{i\omega t}f_\omega(x) + \frac{A}{2}e^{-i\omega t}f_\omega(x)\,,
\end{equation}
where $A \in [0,1]$ is a continuous parameter that may be understood as an amplitude of the perturbation added on top of the Q-ball. We will see that this perturbation is, in fact, a quasi-normal mode (QNM) of the Q-ball. 

%%%%%%%%%%%%%%%%%%%%%%%%%%%%%%%%%%%%%%%%%%%%%%%%
\section{The complex $\phi^6$ model}
\label{model}
%%%%%%%%%%%%%%%%%%%%%%%%%%%%%%%%%%%%%%%%%%%%%%%%

Let us consider a complex $\phi^6$ model in $(1+1)$ dimensions
\cite{Axenides:1999hs,Battye:2000qj,Bowcock,Ciurla:2024ksm}
\begin{equation}
    \mathcal{L} = \partial_\mu\phi\partial^\mu\phi^*-V\left(|\phi|^2\right)\, ,
    \label{lag}
\end{equation}
where the potential of the self-interacting complex scalar field is
\begin{equation}
    V\left(|\phi|^2\right)  =|\phi|^2-|\phi|^4+\beta |\phi|^6\,.
    \label{pot}
\end{equation}
It is known that this theory supports Q-balls. They solve the corresponding equations of motion
\begin{equation}
\phi_{tt}-\phi_{xx}+\frac{\partial V}{\partial |\phi|^2}\phi=0
\label{field_eq}
\end{equation}
provided that $\phi$ has the form (\ref{Qball}) and the profile obeys
\begin{equation}
    \label{stationary}
    \frac{df}{dx}=\pm\sqrt{V(f^2) -\omega^2 f^2}.
\end{equation}
The exact solution reads \cite{Axenides:1999hs,Battye:2000qj,Bowcock}
\begin{equation}
    f_\omega(x) = \frac{\sqrt{2}\epsilon }{\sqrt{1+\sqrt{1-4\beta\epsilon ^2}\cosh(2\epsilon x)}}\, ,
\label{solution}
\end{equation}
where $\epsilon =\sqrt{1-\omega^2}$ is the amplitude of the soliton. The corresponding energy and $U(1)$ charge read
\begin{equation}
    E (\omega) = \frac{4\omega {\epsilon } + Q (4\beta-1 + 4 \beta \omega^2 )}{8\omega \beta}\,.
\label{EQ}
\end{equation}
\begin{equation}
Q(\omega) = \frac{4\omega}{\sqrt \beta}{\rm arctanh}\left(\frac{1-\sqrt{1-4\beta {\epsilon }^2}}{2{\epsilon } \sqrt \beta}\right)\,.
\end{equation}
The Q-balls exist for $\omega \in [\omega_{min}, \omega_{max}]$. The upper bound is just the mass threshold, $\omega_{max}=1$, and the lower bound is $ \omega_{\rm min}(\beta) = \sqrt{1-\frac{1}{4\beta}}$ \cite{Bowcock}. 
In the small amplitude limit, where $\epsilon\to 0$ or equivalently $\omega\to 1$ the profile simplifies to 
\begin{equation}
    f_{\omega\to1}(x) \approx \frac{\sqrt{2}\epsilon}{\sqrt{1+\cosh(2\epsilon x)}}=\frac{\epsilon}{\cosh(\epsilon x)}\,.
\end{equation}

The $\phi^6$ model also supports oscillons. Small-amplitude oscillons can be approximated by (\ref{oscillon}). The leading contribution is 
\begin{equation}
    g_1(x) = \frac{2}{\sqrt{3}}\frac{\epsilon}{\cosh(\epsilon x)}\approx\frac{1.1547\epsilon}{\cosh(\epsilon x)}\,,
\end{equation}
which quite well agrees with the profile of the Q-ball. Therefore, as we claimed before, the small-amplitude oscillon can be approximated as a superposition of two Q-balls of the opposite charge.

%%%%%%%%%%%%%%%%%%%%%%%%%%%%%%%%%%%%%%%%%%%%%%%%
\section{Polarizing the Q-ball} 
\label{polarizing}
%%%%%%%%%%%%%%%%%%%%%%%%%%%%%%%%%%%%%%%%%%%%%%%%

Now we will investigate what happens if we superpose the Q-ball and anti-Q-ball with an arbitrary ratio
\begin{equation}
    \phi(x,t) = \left(1-\frac{A}{2}\right)e^{i\omega t}f_\omega(x) + \frac{A}{2}e^{-i\omega t}f_\omega(x)\,.
    \label{config}
\end{equation}
We call this an arbitrary polarized Q-ball. 
Of course, this configuration is not a solution to the equations of motion, but we can use it as the initial conditions 
\begin{equation}\label{init_cond}
    \phi(x,0)=f(x)\,,\;\;\phi_t(x, 0)=i\omega (1-A)f(x)
\end{equation}
and analyze which solution it relaxes to. 

Note, that the field described by the above formula has a charge equal to
\begin{equation}
    Q(A;\omega) = (1-A)Q(\omega)\,,
\end{equation}
where $Q(\omega)$ is the charge of the Q-ball with frequency $\omega$. This means that the polarized Q-balls have a smaller charge; therefore, one can expect that they could decay to Q-balls with a smaller charge and higher frequency.

% When $A$ is small the configuration (\ref{config}) describes a {\it perturbed} Q-ball. Small, linear perturbations of the Q-ball have recently been studied in \cite{Ciurla:2024ksm}. Q-balls with $\omega$ close to 1 have a single bound mode (BM) and a very long lived quasi-normal mode (QNM). Each of these modes have two components, oscillating with (real) frequencies $\omega\pm \rho_i$. In case of a bound mode, both of the frequencies are inside the mass gap, while in the case of the QNM only one of the components is in the scattering spectrum. Of course, one should remember that the QNM has also an imaginary (damping) part of the frequency, 

When $A$ is small, the configuration (\ref{config}) describes a {\it perturbed} Q-ball. Small, linear perturbations of the Q-ball have recently been studied in \cite{Ciurla:2024ksm}. Q-balls with $\omega$ sufficiently close to 1 have a single bound mode (BM) and a very long-living quasi-normal mode (QNM). Each of these modes has two components that oscillate at frequencies $\omega\pm \rho_i$. In the case of the bound mode, both frequencies are within the mass gap. In the case of the QNM the eigenvalue $\rho$ is complex-valued with a very small imaginary part, responsible for slow damping. The real part of one of the frequencies of the QNM lies within the mass gap and the other in the continuous spectrum. This indicates that the QNM is a Feschbach resonance. 

% The real part of frequency The real parts of frequencwhile in the case of the QNM only one of the components is in the scattering spectrum. Of course, one should remember that the QNM has also an imaginary (damping) part of the frequency, $\Omega_\textrm{q}=\omega_{\textrm{q}}+ i \Gamma_{\textrm{q}}$. Here, $\omega_q=\omega \pm \rho_\textrm{q}$.
\begin{figure}
	\centering
	\includegraphics[width=1\columnwidth]{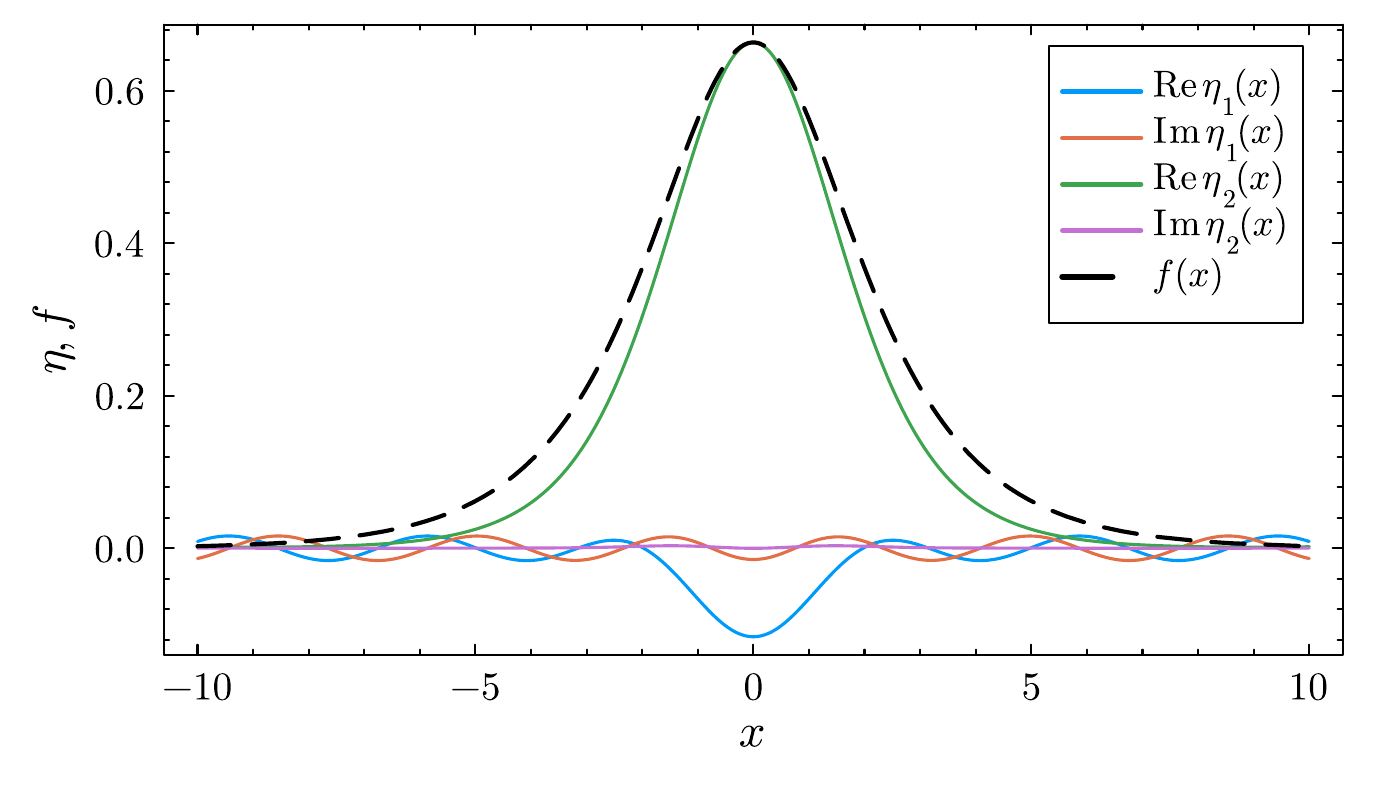}
	\caption{Comparison between the profiles of QNM $\eta_{1,2}$ and a Qball for $\beta=0.26$ and $\omega=0.8$. The frequency of the QNM is  $\omega_{q}=\omega \mp \rho_{q}=0.8 \pm 1.31159$, while the damping part is $\Gamma_{q}=0.00034$. Observe that only $\eta_2$ has a profile coinciding with the profile of the Q-ball.}\label{fig:Comparison}
\end{figure}

Interestingly, for the Q-balls with $\omega$ in a wide range, the profile of the component of the QNM located below the mass threshold, i.e., corresponding to $\omega-\rho_\textrm{q}$, is very similar to the profile of the Q-ball itself, see Fig. \ref{fig:Comparison}.

All this shows that the polarized configuration in the small $A$ limit corresponds to the Q-ball that hosts the BM and the lower frequency component of the QNM. 

Now we evolve the polarized initial configuration for bigger $A$. As $A$ increases, so does the amplitude of the linearized modes. We found that it concerns mostly the amplitude of the QNM. When modes are excited significantly enough, their frequencies change due to nonlinearities. In the Fourier spectrum, higher harmonics are visible as well. 
\begin{figure}
	\centering 
	\includegraphics[width=1\columnwidth]{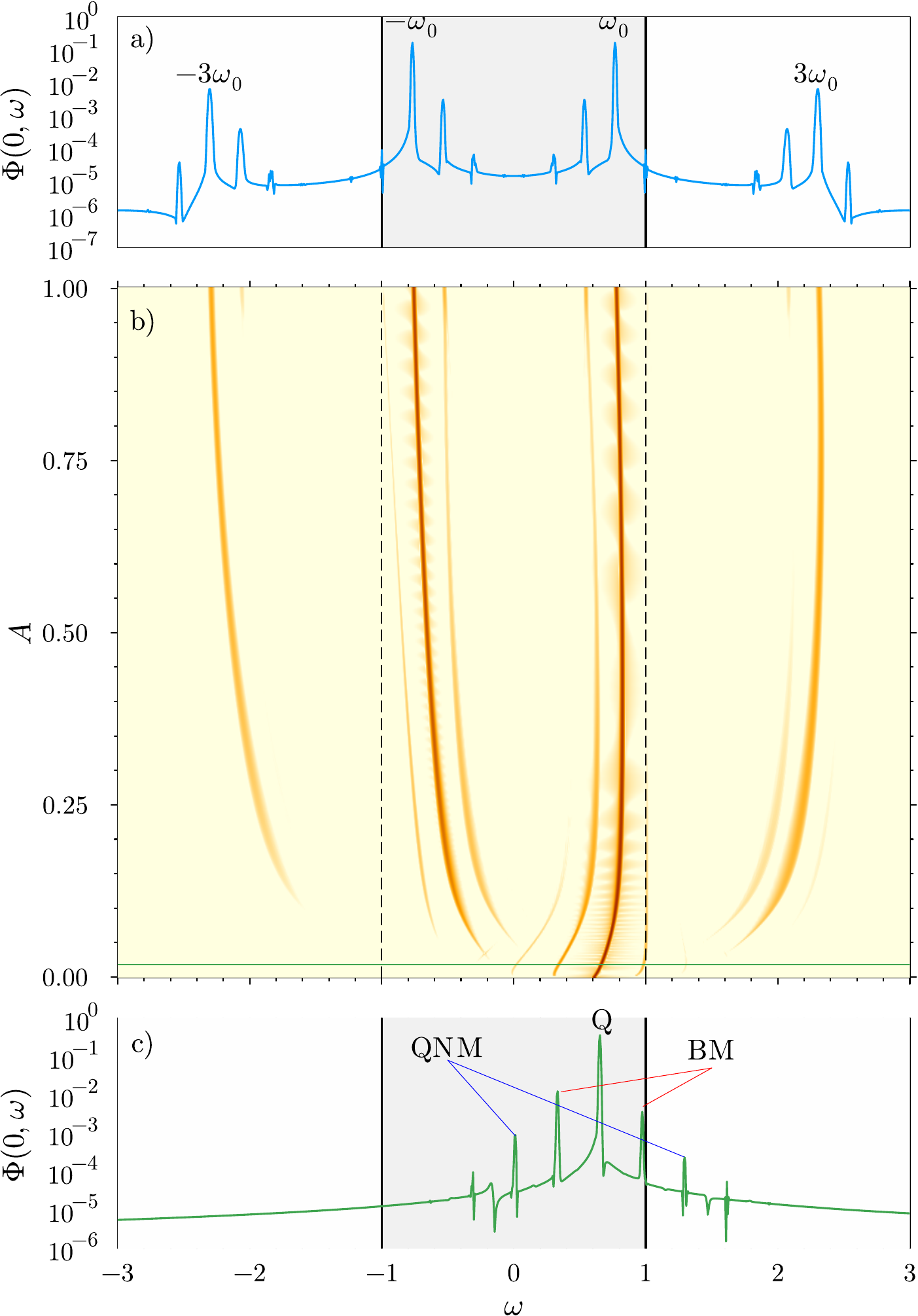}
	\caption{Power spectra for different values of $A$ starting from small amplitude perturbation, $A=0.02$ (c) changing smoothly (b) to an oscillon, A=1, (a).}
	\label{fig:flow}%
\end{figure}

In Fig. \ref{fig:flow} we present the flow of the spectrum for the initial configuration (\ref{config}) with $\omega=0.6$ as we smoothly change $A$ up to $A=1$, i.e., to the oscillon limit. For concreteness, we used $\beta=0.26$.

First of all, one notices that the frequency of the Q-ball, $\omega$, changes into the fundamental frequency of the oscillon, $\omega_O$. This coincides with the fact that approximately the oscillon is a superposition of the Q-ball and anti Q-ball located on top of each other. 

Secondly, also the Q-ball QNM finds its counterparts in the oscillon limit. The frequency of the lower component of the QNM, $\omega-\rho_q$ changes to $-\omega_O$. This is important because the power spectrum of the oscillon is symmetric under $\mathbb{Z}_2$ transformation, $\omega \to - \omega$. The frequency of the higher component, $\omega+\rho_q$, changes into the third harmonic of the fundamental oscillon frequency, $3\omega_O$, which is responsible for the main radiation channel leading to the oscillon decay. 

Thirdly, the resulting oscillon carries some modulations. This is the fundamental property of a generic oscillon that its amplitude oscillates as well. Hence, a second frequency is involved. It has recently been advocated that such a modulated oscillon is in fact a bound state of two fundamental (non-modulated) oscillons, each with its own frequency $\omega_{O_{1,2}}$ \cite{Blaschke:2024uec}. Then the modulation frequency is just $\omega_{mod}=|\omega_{O_1}-\omega_{O_2}|$. Here, the first fundamental oscillon is just the one built out of a Q-ball and an anti-Q-ball. Its frequency is $\omega_O$. The second frequency (identified with another fundamental oscillon forming the bound state) is visible in the power spectrum as two peaks that emerge from the BM of the Q-ball with $\omega_0 \pm \rho_{\textrm{b}}$. Thus, this additional fundamental oscillon arises from the linear BM of the Q-ball. We note that one of these peaks is located very close to the mass threshold. This close proximity to the mass threshold oscillon has recently been identified as a seed of amplitude modulation \cite{Blaschke:2024uec}. We also note that the idea that an oscillon may arise smoothly from a normal mode was elaborated in \cite{Romanczukiewicz:2017gxb}. 

%It is important to notice that the additional two peaks in the oscillon power spectrum, which we associate with the modulations come from the BM of the Q-ball. The upper frequency component, $\omega+\rho_\textrm{b}$, approaches very close to the mass threshold as $A\to 1$, $\omega_+$. Such a close to the mass threshold mode has been recently identified as a seed of the amplitude modulation and interpreted as the frequency of the second oscillon which forms the two-oscillon bound state. Indeed, the frequency of the modulations is $\omega_{mod}=\omega_+-\omega_O$ \cite{Blaschke:2024uec}. Next, the lower frequency component, $\omega-\rho_\textrm{b}$ transmutes into the second peak in the oscillon spectrum, $\omega_-=\omega_O-\omega_{mod}$. 
 \begin{figure}
	\centering
	\includegraphics[width=1\columnwidth]{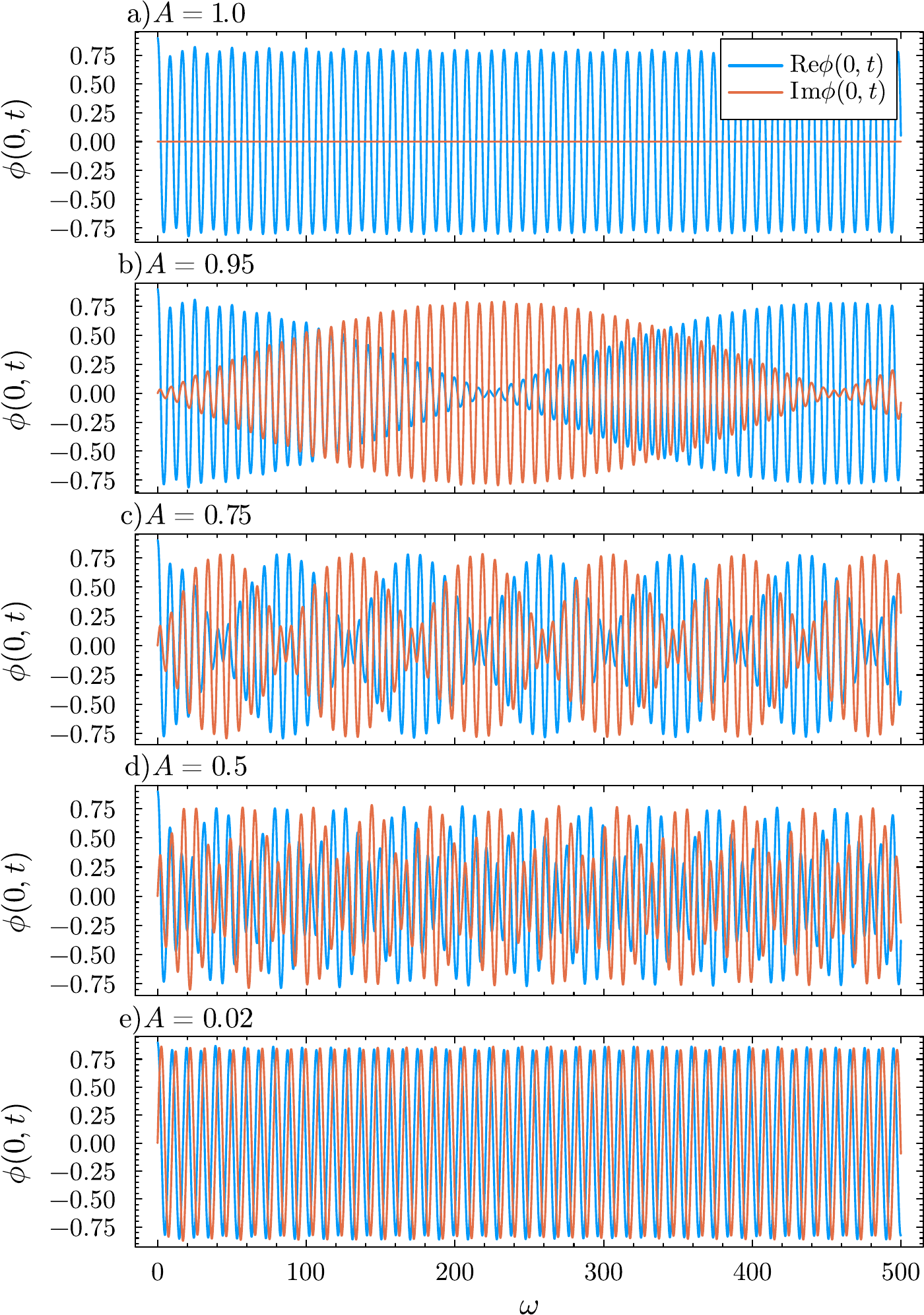}
	\caption{Time evolution of the real and imaginary component of the complex field $\phi$ for different $A$. Here $\beta=0.26$ and $\omega=0.6$}\label{fig:oscillon_Qball}%
\end{figure}

There is also an interesting regime where $A$ is close to 1. It leads to a complex oscillon-like state, for which two main frequencies are not $\mathbb{Z}_2$ symmetric. Hence, we have two peaks at $\omega_O$ and $\omega_{\tilde{O}}$ such that $\omega_O \neq - \omega_{\tilde{O}}$. In Fig. \ref{fig:oscillon_Qball}, we show an example of such an oscillating solution for $A=0.95$. %It looks like two oscillons, each in each component of the complex field, which are modulated with a small frequency $\omega_O+\omega_{\tilde{O}}\approx  0.7656 -0.7526 = 0.013$. This is a modulation related to the coupling between the Re $\phi$ and Im $\phi$ components. This picture holds even for the parameter $A$ much below 1, i.e., $A=0.75$. All the time the modulation of these oscillons is shifted in phase by half of the period.
In the appendix, we modify the usual perturbative expansion and obtain an approximation of these complex oscillon-like solutions. 

 \vspace*{0.2cm}
 
To summarize, we clearly see that not only the oscillon itself arises from the Q-ball but, even more exciting, also the spectral properties of the oscillon are inherited from the modes of the Q-ball. 

%%%%%%%%%%%%%%%%%%%%%%%%%%%%%%%%%%%%%%%%%%%%%%%%
\section{Stability}
\label{stability}
%%%%%%%%%%%%%%%%%%%%%%%%%%%%%%%%%%%%%%%%%%%%%%%%
 \begin{figure}
	\centering
	\includegraphics[width=1\columnwidth]{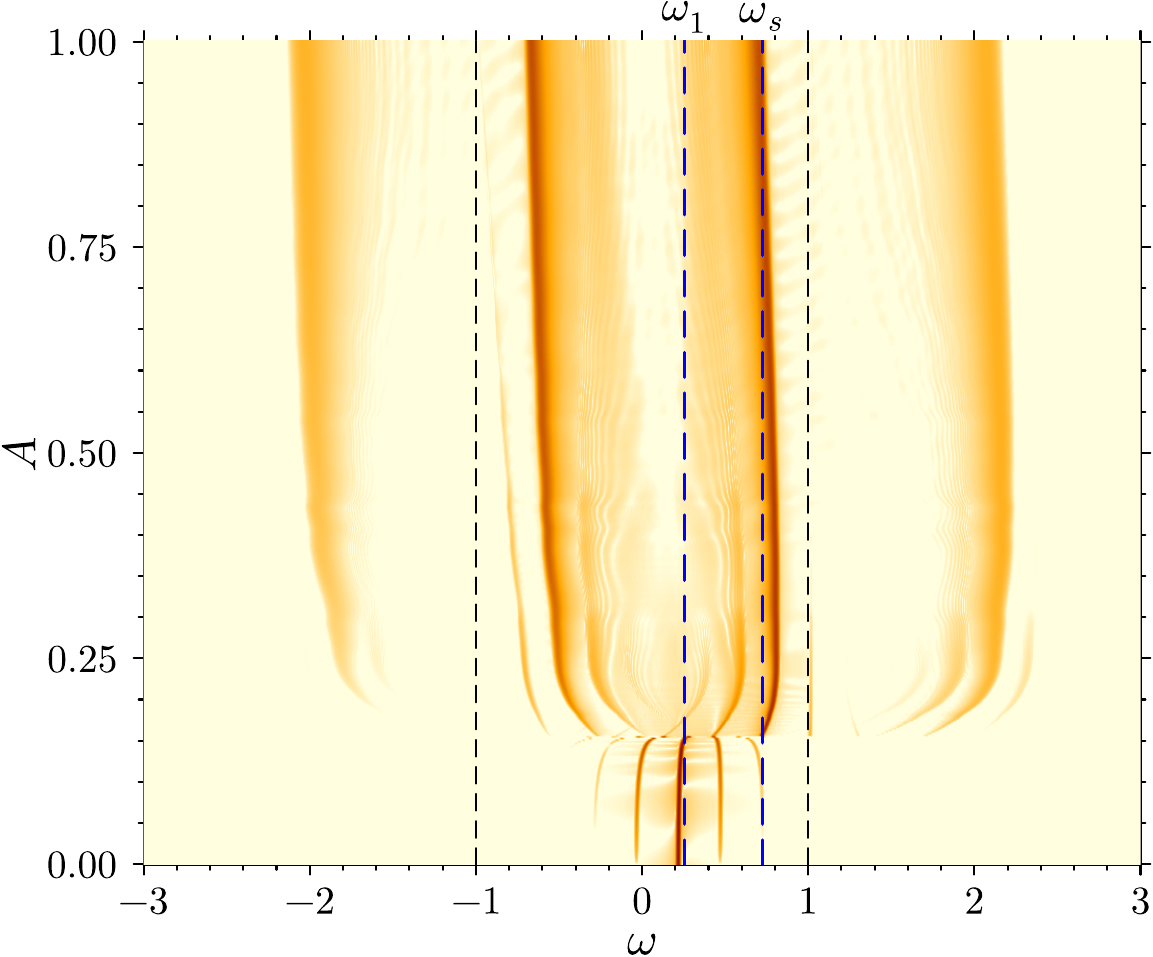}
	\caption{Power spectra for different values of $A$ starting from small amplitude perturbation for $\beta=0.26$ and $\omega=0.2$.}\label{fig:unstable_flow}%
\end{figure}

The Q-balls usually exist only for certain values of frequency $\omega$. The stability condition is typically $dQ/d\omega<0$ \cite{Friedberg:1976me} (although it can be significantly modified in theories with a more complicated target space \cite{Alonso-Izquierdo:2023xni}, \cite{Alonso-Izquierdo:2024ohc}). For the $\phi^6$ theory with $\beta=0.26$ the stability region consists of two segments $\omega \in(\omega_\textrm{min}, \omega_1)\cup (\omega_2,1)$, where
$\omega_\textrm{min} = 0.196116, \omega_1= 0.255100$ and $\omega_2=0.51969$. A Q-ball with frequency in the range $\omega\in(\omega_1, \omega_2)$ is unstable and rapidly decays to a Q-ball with $\omega_s$ within the stable range $(\omega_2, 1)$. The frequency $\omega_s$ is such that the $U(1)$ charge of the initial (unstable) and final (stable) Q-ball is the same, $Q(\omega)=Q(\omega_s)$. 

As we have seen above, the perturbed Q-ball relaxes to a Q-ball with a higher frequency. Thus, the polarized initial condition can change the stable Q-ball into an unstable one. This is visible in Figure \ref{fig:unstable_flow}, where we plot the flow of the power spectrum with $A$. Here $\omega=0.2$ and $\beta=0.26$. At a certain value of the perturbation parameter $A$ the frequency of the perturbed Q-ball reaches $\omega_1$ and enters the unstable regime. It transmutes into the stable excited Q-ball with frequency $\omega_s=0.724441$. In consequence, it is impossible to reach a low-frequency oscillon from the polarized initial conditions. 

In fact, this process resembles the rapid decay of low-frequency oscillons. It is quite common that low-frequency oscillons are quite volatile and quickly radiate some energy before entering the more stable stage, in which they very slowly decay. 

%%%%%%%%%%%%%%%%%%%%%%%%%%%%%%%%%%%%%%%%%%%%%%%%
\section{Conclusions}
\label{conclusions}
%%%%%%%%%%%%%%%%%%%%%%%%%%%%%%%%%%%%%%%%%%%%%%%%

We showed that the oscillons and Q-balls in the complex $\phi^6$ model are approximately "dual" objects. One can be approximately expressed by the other. Namely, a Q-ball can be viewed as a superposition of two oscillons, or equivalently, the oscillon can be treated as a Q-ball and anti-Q-ball pair. In a sense, both represent different types of "polarization" of the complex field. 

Furthermore, the structures of their modes visible in the power spectrum are related. The fundamental frequency $\omega$ of the Q-ball and the lower component of its quasinormal mode are sources of the fundamental frequencies of the oscillon $\pm \omega_O$. The higher component of the quasinormal mode transforms into the radiating harmonic of the oscillon, which is responsible for its decay. 
The bound mode gives rise to the amplitude modulation being a seed for the second small amplitude oscillon that forms a bound state with the main oscillon (this with frequency $\omega_O$). Thus, the dynamical properties of the oscillons are consequences of the dynamical properties of the Q-balls. 

In addition, we found evidence that the decaying of the unstable Q-balls shares some properties of the decay of the oscillons. 

We have also found that, considered here, initial conditions give rise to a whole family of rather stable solutions which, at least in the limit $A \to 1$, can be treated as a complex-valued oscillon with two, not exactly opposite frequencies and nonvanishing total charge. 

%an excited oscillon with a $U(1)$ charge wave. 

The observed relation between the oscillons and Q-balls in the complex model $\phi^6$, and especially a smooth transition between these excitations, may be helpful in a better understanding of the charge-swapping phenomenon \cite{Copeland:2014qra}. This is a long-lived state formed by a spatially separated pair of Q-ball and anti-Q-ball with zero net charge, where the local distribution of the charge changes in a periodic manner. Eventually, such a charge swapping state decays into an oscillon \cite{Xie:2021glp}.

$Q$-balls as well as oscillons appear in various cosmological scenarios. They can be produced in the early universe, especially at the end of inflation \cite{Gleiser:2011xj, Amin:2011hj, Zhou:2013tsa, Aurrekoetxea:2023jwd}. They may contribute to the dark matter sector \cite{Kusenko:2001vu, Amin:2011hu}. They can also play a role in baryogenesis \cite{Enqvist:1997si, Lozanov:2014zfa}. It would be very interesting to understand the physical consequence of the fact that $Q$ balls and oscillons seem to be smoothly deformable one into the other.
%%%%%%%%%%%%%%%%%%%%%%%%%%%%%%%%%%%%%
\section{Acknowledgments}
%%%%%%%%%%%%%%%%%%%%%%%%%%%%%%%%%%%%%
FB acknowledges the institutional support of the Research Centre for Theoretical Physics and Astrophysics, Institute of Physics, Silesian University in Opava.
FB has been supported by the grant no. SGS/24/2024 Astrophysical processes in strong gravitational and electromagnetic fields of compact object.
KS acknowledges financial support from the Polish National Science Center
(grant NCN 2021/43/D/ST2/01122). AW acknowledges support from the Spanish Ministerio de Ciencia e Innovacion (MCIN) 
with funding from the European Union NextGenerationEU (Grant No. PRTRC17.I1) and the Consejeria de Educacion from JCyL through the QCAYLE project, as well as the MCIN Project 1114 No. PID2020-113406GB-I0 and the grant PID2023-148409NB-I00 MTM. 
%%%%%%%%%%%%%%%%%%%%%%%%%%%%%%%%%%%%%%%%%%%%%%%%
\appendix
\section{Complex oscillon expansion}
%%%%%%%%%%%%%%%%%%%%%%%%%%%%%%%%%%%%%%%%%%%%%%%%

The standard approach to obtain oscillon profiles is to use Fourier expansion  (\ref{oscillon}) 
\cite{Fodor:2008du} introducing a small perturbation parameter $\epsilon$ which, at appropriate powers, multiplies the profile function $g_k$. The next step is to introduce new, scaled space and time variables and solve the problem order by order. Every second must deal with resonant terms. Conditions for canceling these terms lead to nonlinear equations for the profiles. We will adopt this method, modify it, and apply it to small-amplitude oscillons in our model. Spectral analysis (Figure \ref{fig:flow}) reveals that for $A=1$ there are two main frequencies $\pm \omega_O$, but for $A$ slightly below 1, the frequencies shift and do not have the same modulus. Therefore, any scheme that describes the profiles must have two frequencies. This problem is more complicated than for standard oscillons, therefore we will limit our considerations only to the lowest orders of the perturbation series.  

Assuming that the field can be approximated with 
\begin{equation}
    \phi(x,t)\approx \sum_{n=-\infty}^{\infty}\epsilon^{|m|+|n|}g_{nm}(x)z_1^nz_2^m\,,
\end{equation}
where
\begin{equation}
    z_1=\exp(i\omega_{O,1}t)\,,\qquad z_2=\exp(-i\omega_{O,2}t)
\end{equation}
and
\begin{equation}
    \omega_{O,i}=1+\sum_{n=1}^{\infty}\epsilon^{n}\varpi_{i,n}\,.
\end{equation}
We will also assume that functions $g_{nm}(x)$ are real-valued. This assumption is not necessary, in general, but is consistent with our initial conditions. However, the scheme in the following can be easily generalized to complex-valued functions.

Following \cite{Fodor:2008du} we introduce new variables
\begin{equation}
    \zeta=\epsilon x\,,\qquad \tau=\sqrt{1-\epsilon^2}t\,.
\end{equation}
Equations of motion now take the following form
\begin{equation}
    (1-\epsilon^2)\phi_{\tau\tau}-\epsilon^2\phi_{\zeta\zeta}+(1-2|\phi|^2+3\beta|\phi|^4)\phi=0\,.
\end{equation}
The first two equations in the expansion are purely algebraic. Here 
\begin{equation}
    g_{00}(x)\equiv0\,,
\end{equation}
which is follows from the fact that $\phi=0$ is a symmetric vacuum.
In the first order in $\epsilon$ we find a set of two equations
\begin{equation}
    (-\omega_{O,1}^2+1)g_{1, 0}=0\qquad (-\omega_{O,2}^2+1)g_{0, 1}=0\,.
\end{equation}
As $\omega_{O,i}=1+\mathcal{O}(\epsilon)$, we find that $g_{1,0}(x)$ and $g_{0,1}(x)$ could be arbitrary functions. They will be determined by the higher order terms, and we will assume they are not equivalently equal 0.
\begin{figure}
	\centering
	\includegraphics[width=1\columnwidth]{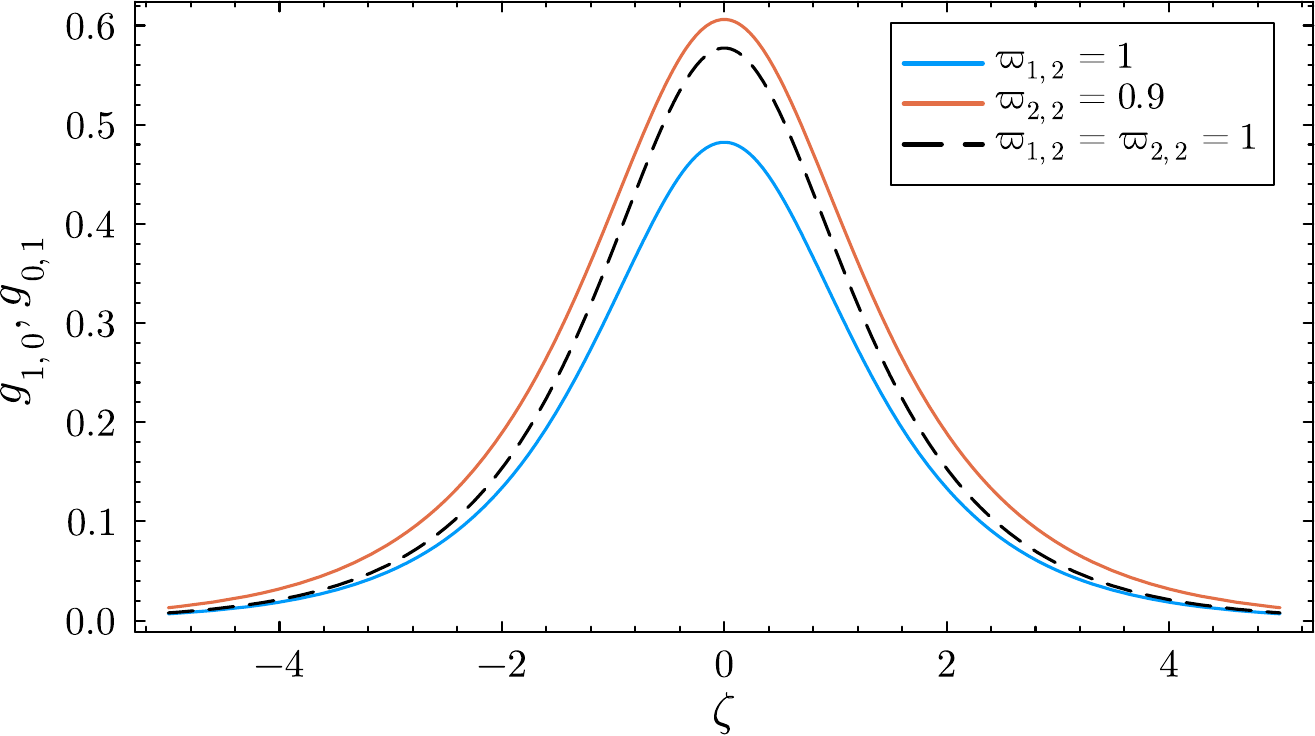}
	\caption{Solutions $g_{0,1}$ and $g_{1,0}$ to (\ref{system1})-(\ref{system2}) for a single frequency standard oscillon (dashed line) and for a two-frequency oscillon (solid lines).}\label{fig:oscillon_profiles}%
\end{figure}

The second order equation takes the form:
\begin{equation}
    \begin{aligned}
        &-3g_{2,0}z_1^2-3g_{0,2}z_2^2+g_{1,1}z_1z_2-\\ 
        &-2\varpi_{1,1}g_{1,0}z_1-2\varpi_{2,1}g_{0,1}z_2-\\
        &-2\varpi_{1,1}g_{-1,0}z_1^{-1}-2\varpi_{2,1}g_{0,-1}z_2^{-1}-\\
        &-3g_{-2,0}z_1^{-2}-3g_{0,-2}z_2^{-2}+g_{-1,-1}z_1^{-1}z_2^{-1}-\\
        &-3g_{1,-1}z_1z_2^{-1}-3g_{-1,1}z_1^{-1}z_2=0.        
    \end{aligned}
\end{equation}
Since neither $g_{1,0}$ nor $g_{0,1}$ is equal to 0, $\varpi_{2,1}=\varpi_{1,1}=0$. This leaves freedom to choose $g_{-1,0}$ and $g_{0,-1}$ as arbitrary functions. 
Setting them to zero is consistent with the spectrum of the oscillon. Other functions must vanish 
\begin{equation}
    g_{2,0}=g_{0,2}=g_{1,-1}=g_{-1,1}=g_{-2,0}=g_{0,-2}=g_{-1,-1}=0\,.
\end{equation}

In $\mathcal{O}(\epsilon^3)$ the equation is no longer algebraic, but contains also derivative terms proportional to $z_1$ and $z_2$.  The demand that these terms vanish leads to a coupled system of ODEs for functions $g_{1,0}$ and $g_{0,1}$

\begin{eqnarray}
    g_{1,0}''&=&(1+2\varpi_{1, 2})g_{1,0}-2g_{1,0}^3-4g_{1,0}g_{0,1}^2 \label{system1}\,,\\
    g_{0,1}''&=&(1+2\varpi_{2, 2})g_{0,1}-2g_{0,1}^3-4g_{0,1}g_{1,0}^2\label{system2}\,,    
\end{eqnarray}
where $'=d/d\zeta$. This condition is equivalent to cancelling resonance terms described in \cite{Forgacs:2008az}.
The above system contains arbitrary (but small) corrections to frequencies $\varpi_{1, 2}$ and $\varpi_{2, 2}$. This is the key difference 

A similar system of equations can be derived for $g_{-1,0}$ and $g_{0,-1}$ (terms proportional to $z_{1,2}^{-1}$ but the setting $g_{-1,0}=g_{0,-1}\equiv0$, which is a solution, is consistent with our initial conditions. 
% \section{Integrability}
% \trom{copied from us}\\
% The next step is to observe that, around the vacuum, the universal RG $Q$-ball equation can be approximated by the {\it integrable complex sine-Gordon} ($\mathbb{C}$sG) equation:
% \begin{equation}
% \bigl(\partial^2 +1\bigr)\Psi = \Psi |\Psi|^2 - \bar\Psi \frac{\partial_\mu \Psi \partial^\mu \Psi}{1- |\Psi|^2}\,.
% \end{equation} 
% The sense, in which this equation is ``close'' to RG equation \ref{} is that it has exactly the same $Q$-ball solution given in Eq.~\ref{}. Furthermore, the Lagrangian describing $\mathbb{C}$sG model differs from \ref{} by a multiplicative factor of the form $1+ \mathcal{O}\bigl(|\Psi|^2\bigr)$, i.e. 
% \begin{equation}
% \mathcal{L}_{\mbox{$\mathbb{C}$sG}} = \frac{1}{1-|\Psi|^2}\Bigl(\partial_\mu \bar\Psi \partial^\mu \Psi - |\Psi |^2+|\Psi |^4\Bigr)\,.
% \end{equation}

% Importantly, due to the integrability the $\mathbb{C}$sG model supports a two Q-ball solution \cite{Bowcock:2008dn} describing a motion of two single $Q$-balls, each with its own scale parameter:\\
% % \begin{strip}
% % \vspace*{1cm}
% \begin{equation}
% \Psi_{12} = \frac{i \bigl(\omega_1 -\omega_2\bigr)\Bigl(\frac{\lambda_1}{\cosh(\lambda_1 x)}e^{i \omega_1 t}-\frac{\lambda_2}{\cosh(\lambda_2 x)}e^{i \omega_2 t}\Bigr)}
% {1-\omega_1\omega_2 -\lambda_1\lambda_2 \Bigl(\tanh(\lambda_1 x) \tanh(\lambda_2 x)+\frac{\cos(t(\omega_1-\omega_2))}{\cosh(\lambda_1 x)\cosh(\lambda_2 x)}\Bigr)}
% \end{equation}
% %\end{strip}
% \begin{equation}
%      \omega_{1,2}\equiv \sqrt{1-\lambda_{1,2}^2} \,.
% \end{equation}

\bibliographystyle{JHEP}
\bibliography{QB_QNM_O_v1}
\end{document}